\documentclass[aps,prd,onecolumn,showpacs,showkeys,amsmath,amssymb]{revtex4}
\usepackage{epsfig} 
\usepackage{amsmath}  
\usepackage{makecell}
\usepackage{graphicx}
\usepackage{array}
\usepackage{subfigure}
\usepackage{bm}
\usepackage{longtable}
\usepackage{tabularx}
\usepackage{booktabs}
\usepackage{appendix}
\usepackage{caption2}
\usepackage{amsfonts}
\usepackage{amsmath}
\usepackage{graphicx}
\usepackage{subfigure}
\usepackage{dcolumn}
\usepackage{siunitx}
\usepackage{bm}
\usepackage{booktabs}
\usepackage[utf8]{inputenc}

\graphicspath{{figs/}}
\usepackage{float}
\usepackage{longtable,lscape}
\usepackage{txfonts}
\usepackage{overpic}
\usepackage{amssymb}
\usepackage{indentfirst}
\usepackage{epsfig}
\usepackage{feynmf}   
\usepackage{epstopdf}   
\usepackage{slashed}  
\usepackage{multirow}

\usepackage[colorlinks, citecolor=blue,anchorcolor=red,menucolor=red, linkcolor=red,filecolor=red,runcolor=red,urlcolor=blue,frenchlinks=red]{hyperref}

\usepackage{ulem}
\usepackage{color}

\makeatletter
\newcommand{\figcaption}{\def\@captype{figure}\caption}
\newcommand{\tabcaption}{\def\@captype{table}\caption}

\newcommand{\Rmnum}[1]{\expandafter\@slowromancap\romannumeral #1@}

\def\hlinewd#1{%
  \noalign{\ifnum0=`}\fi\hrule \@height #1 \futurelet
   \reserved@a\@xhline}
\makeatother

\begin{document}

\title{ Mass spectra for the $cc\bar{b}\bar{b}$/$bb\bar{c}\bar{c}$ tetraquark states}

\author{Qi-Nan Wang}
\author{Zi-Heng Yang}
\author{Wei Chen$^1$}
\email{chenwei29@mail.sysu.edu.cn}
\author{Hua-Xing Chen$^2$}
\email{hxchen@seu.edu.cn}
\affiliation{$^1$School of Physics, Sun Yat-Sen University, Guangzhou 510275, China
\\
 $^2$School of Physics, Southeast University, Nanjing 210094, China }

\begin{abstract}
We have studied the masse spectra for the $cc\bar{b}\bar{b}$/$bb\bar{c}\bar{c}$ tetraquark states  with quantum numbers $J^{P}=0^{\pm},1^{\pm}$, and $2^{+}$. We systematically construct the interpolating currents with various spin-parity quantum numbers and calculate their two-point correlation functions in the framework of QCD moment sum rule method. Our calculations show that the masses are about $12.3-12.4$ GeV for the positive parity $cc\bar{b}\bar{b}$ tetraquark ground states with $J^{P}=0^+, 1^+, 2^+$, while $12.8-13.1$ GeV for the negative parity channels with $J^{P}=0^-, 1^-$. The mass predictions 
for the positive parity $cc\bar{b}\bar{b}$ ground states are lower than the $B_{c}B_{c}$ threshold, implying that these tetraquarks can only decay via weak interaction and thus are expected to be stable and narrow. 

\end{abstract}


\pacs{12.39.Mk, 12.38.Lg, 14.40.Ev, 14.40.Rt}
\keywords{Di-$J/\psi$, Exotic state, QCD sum rules}
\maketitle

\section{Introduction}
The existence of multiquarks ($qq\bar{q}\bar{q}$ tetraquark, $qqqq\bar q$ pentaquark, etc.) was proposed firstly by Murray Gell-Mann and George Zweig at the birth of quark model in 1964~\cite{GellMann:1964nj,zweigSU(3)}, in which hadrons were classified as $q\bar q$ mesons and $qqq$ baryons. Later in 1977, Jaffe applied the diquark-antidiquark genuin tetraquark configuration to the light scalar mesons, and successfully explained their mass ordering problem and the decay property of $f_0(980)$ ~\cite{1977-Jaffe-p267-267,1977-Jaffe-p281-281}. However, another interesting configuration of hadron molecule was also proposed to interpret the $f_0(980)$ as a $K\bar K$ bound state~\cite{1990-Weinstein-p2236-2236}. Actually, the compact diquark-antidiquark genuin tetraquark and loosely bound hadron molecules are two important but distinct pictures to understand the underlying structures of exotic hadrons, especially after the observations of numerous XYZ states and hidden-charm pentaquark states~\cite{2016-Chen-p1-121,2017-Ali-p123-198,2017-Lebed-p143-194,2018-Guo-p15004-15004,2019-Liu-p237-320,2020-Brambilla-p1-154}.

In 2017, the CMS Collaboration reported their measurement of an exotic excess around $18.4$ GeV in the four lepton channel with a global significance 
of 3.6$\sigma$~\cite{Khachatryan:2016ydm}, which had inspired lots of theoretical studies on the $bb\bar b\bar b$  four-bottom tetraquark states~\cite{Chen:2016jxd,Anwar:2017toa,Esposito:2018cwh,Hughes:2017xie,Karliner:2016zzc,Wu:2016vtq,Richard:2017vry,Bai:2016int,Chen:2019dvd,Debastiani:2017msn}. However, such an exotic structure was not confirmed by the later experiments~\cite{2018-Aaij-p86-86,2020-Sirunyan-p135578-135578}. Very recently, the LHCb Collaboration announced the evidence for new resonance structures in the di-$J/\psi$ mass spectrum\cite{Aaij:2020fnh}, in which a narrow structure $X(6900)$ around 6.9 GeV in addition to a broad structure range from $6.2-6.8$ GeV were discovered. The observation of these new structures has immediately inspired widespread research interest~\cite{2020-Chen-p1994-2000,Albuquerque:2020hio,An:2020jix,Bai:2016int,Chen:2019dvd,Chen:2020xwe,Debastiani:2017msn,Feng:2020riv,Giron:2020wpx,Gordillo:2020sgc,Guo:2020pvt,Huang:2021vtb,Jin:2020jfc,Karliner:2020dta,Ke:2021iyh,Li:2019uch,Li:2021ygk,Liang:2021fzr,Liu:2019zuc,liu:2020eha,Lu:2020cns,Ma:2020kwb,Pal:2021gkr,Sonnenschein:2020nwn,Wan:2020fsk,Wang:2018poa,Wang:2019rdo,Wang:2020gmd,Wang:2020ols,Wang:2020tpt,Wang:2021kfv,Weng:2020jao,Yang:2020rih,Yang:2020wkh,Zhang:2020xtb,Zhao:2020cfi,Zhao:2020nwy,Zhu:2020snb,Zhu:2020xni,Gong:2020bmg,Cao:2020gul,Yang:2021zrc}. For the inside structure of $X(6900)$, the compact diquark-antidiquark tetraquark configuration shall be favored rather than  
the hadron molecule configuration since there are no color singlet light mesons that may be exchanged between two charmonia to produce binding interactions~\cite{2020-Maiani-p-,2020-Chao-p1952-1953}. 

Comparing to the $cc\bar{c}\bar{c}$ and $bb\bar{b}\bar{b}$ systems, the $cc\bar b\bar b$ tetraquark states are very interesting since they are doubly-charged and have no annihilation decay channels. They are expected to be stable if they lie below the $2B_c$ threshold. In Ref.~\cite{Anwar:2017toa}, the ground state ,energy of the $cc\bar b\bar b$ tetraquarks were calculated in a nonrelativistic effective field theory and in a diquark model. They gave an upper limit on mass of $cc\bar{b}\bar{b}$ tetraquark as 12.58 GeV below the $2B_{c}$ threshold, indicating the possibility of stable $cc\bar{b}\bar{b}$ tetraquark against strong decays. However, such possibility was not supported by the investigations of the approach of chromomagnetic interaction (CMI) model~\cite{SilvestreBrac:1992mv,Wu:2016vtq} and the constituent quark model~\cite{Czarnecki:2017vco}, where no bound state of the $cc\bar{b}\bar{b}$ tetraquark was found. In Ref.~\cite{Liu:2019zuc}, the authors studied the $cc\bar{b}\bar{b}$ tetraquark with $J^{PC}=(0,1,2)^{+}$ in the quark potential model and obtained the masses about 12.9-13.0~GeV, which is above $B_{c}^{(*)}B_{c}^{(*)}$ threshold. More theoretical studies are needed to investigate the stability of the $cc\bar b\bar b$ tetraquark systems. In this work, we shall systematically study the mass spectra of $cc\bar{b}\bar{b}$ tetraquark states with quantum numbers $J^{P}=0^{\pm},1^{\pm}$, and $2^{+}$ by using the method of QCD moment sum rules~\cite{Shifman:1978bx,Shifman:1978by,Reinders:1984sr}.

This paper is organized as follows. In Sec.~\Rmnum{2}, we construct the interpolating currents of the  $cc\bar{b}\bar{b}$ tetraquark  systems with $J^{P}=0^{\pm},1^{\pm}$, and $2^{+}$, respectively. In Sec.~\Rmnum{3}, we evaluate the correlation functions for these interpolating currents. We extract the masses for the tetraquark states by performing the QCD moment sum rule analyses in Sec.~\Rmnum{4}. The last section is a brief summary and discussion.

\section{Interpolating tetraquark currents }
In this section, we construct the interpolating currents of the  $cc\bar{b}\bar{b}$ tetraquark  systems with $J^{P}=0^{\pm},1^{\pm}$, and $2^{+}$, respectively. There are five independent diquark fields,
$q_{a}^{T} C \gamma_{5} q_{b},~ q_{a}^{T} C q_{b},~ q_{a}^{T} C \gamma_{\mu} \gamma_{5} q_{b},~ q_{a}^{T} C \gamma_{\mu} q_{b},\text{and} ~q_{a}^{T} C \sigma_{\mu \nu} q_{b}$, where $q$ is the quark field, $a,b$ represent the color indices, $C$ is the charge conjugate operator, and $T$ stands for the transpose of the quark fields.
The $q_{a}^{T} C \gamma_{5} q_{b}$ ($J^{P}=0^{+}$) and $q_{a}^{T} C \gamma_{\mu} q_{b}$ ($J^{P}=1^{+}$) are $S$-wave operators while $ q_{a}^{T} C q_{b}$ ($J^{P}=0^{-}$) and $q_{a}^{T} C \gamma_{\mu} \gamma_{5} q_{b}$ ($J^{P}=1^{-}$) are $P$-wave operators.
The $q_{a}^{T} C \sigma_{\mu \nu} q_{b}$ contains both $S$-wave and $P$-wave pieces according to its different components. We can obtain the $cc\bar{b}\bar{b}$ tetraquark interpolating currents with various quantum numbers via the combinations of these diquark and antidiquark fields.
\begin{itemize}
	\item The tetraquark interpolating currents with $J^{P}=0^{-}$ are
\begin{equation}
\begin{aligned}
\eta_{1}^{-}&=c_{a}^{T} C c_{b}\left(\bar{b}_{a} \gamma_{5} C \bar{b}_{b}^{T}+\bar{b}_{b} \gamma_{5} C \bar{b}_{a}^{T}\right) \, ,\\
\eta_{2}^{-}&=c_{a}^{T} C \gamma_{5} c_{b}\left(\bar{b}_{a} C \bar{b}_{b}^{T}+\bar{b}_{b} C \bar{b}_{a}^{T}\right)\, , \\
\eta_{3}^{-}&=c_{a}^{T} C \sigma_{\mu \nu} c_{b}\left(\bar{b}_{a} \sigma^{\mu \nu} \gamma_{5} C \bar{b}_{b}^{T}-\bar{b}_{a} \sigma^{\mu \nu} \gamma_{5} C \bar{b}_{a}^{T}\right)
\end{aligned}
\label{J0m}
\end{equation}
in which $\eta_{1}^{-}$ and $\eta_{2}^{-}$ are in symmetric color structure $[6_{c}]_{cc}\otimes[\bar{6}_{c}]_{\bar{b}\bar{b}}$ while $\eta_{3}^{-}$ in antisymmetric color structure $[\bar{3}_{c}]_{cc}\otimes[3_{c}]_{\bar{b}\bar{b}}$.

\item The tetraquark interpolating currents with $J^{P}=0^{+}$ are
\begin{equation}
\begin{aligned}
\eta_{1}^{+} &=c_{a}^{T} C c_{b}\left(\bar{b}_{a} C \bar{b}_{b}^{T}+\bar{b}_{b} C \bar{b}_{a}^{T}\right) \\
\eta_{2}^{+} &=c_{a}^{T} C \gamma_{5} c_{b}\left(\bar{b}_{a} \gamma_{5} C \bar{b}_{b}^{T}+\bar{b}_{b} \gamma_{5} C \bar{b}_{a}^{T}\right) \\
\eta_{3}^{+} &=c_{a}^{T} C \gamma_{\mu} c_{b}\left(\bar{b}_{a} \gamma^{\mu} C \bar{b}_{b}^{T}-\bar{b}_{b} \gamma^{\mu} C \bar{b}_{a}^{T}\right) \\
\eta_{4}^{+} &=c_{a}^{T} C \gamma_{\mu} \gamma_{5} c_{b}\left(\bar{b}_{a} \gamma^{\mu} \gamma_{5} C \bar{b}_{b}^{T}+\bar{b}_{b} \gamma^{\mu} \gamma_{5} C \bar{b}_{a}^{T}\right) \\
\eta_{5}^{+} &=c_{a}^{T} C \sigma_{\mu \nu} c_{b}\left(\bar{b}_{a} \sigma^{\mu \nu} C \bar{b}_{b}^{T}-\bar{b}_{b} \sigma^{\mu \nu} C \bar{b}_{a}^{T}\right)
\end{aligned}
\label{J0p}
\end{equation}
in which $\eta_{1}^{+}$, $\eta_{2}^{+}$ and $\eta_{4}^{+}$ are in symmetric color structure $[6_{c}]_{cc}\otimes[\bar{6}_{c}]_{\bar{b}\bar{b}}$ while $\eta_{3}^{+}$ and $\eta_{5}^{+}$ in antisymmetric color structure $[\bar{3}_{c}]_{cc}\otimes[3_{c}]_{\bar{b}\bar{b}}$.

\item The tetraquark interpolating currents with $J^{P}=1^{-}$ are
\begin{equation}
\begin{aligned}
\eta_{1 \mu}^{-} &=c_{a}^{T} C \gamma_{\mu} \gamma_{5} c_{b}\left(\bar{b}_{a} \gamma_{5} C \bar{b}_{b}^{T}+\bar{b}_{b} \gamma_{5} C \bar{b}_{a}^{T}\right) \\
\eta_{2 \mu}^{-} &=c_{a}^{T} C \gamma_{5} c_{b}\left(\bar{b}_{a} \gamma_{\mu} \gamma_{5} C \bar{b}_{b}^{T}+\bar{b}_{b} \gamma_{\mu} \gamma_{5} C \bar{b}_{a}^{T}\right) \\
\eta_{3 \mu}^{-} &=c_{a}^{T} C \sigma_{\mu \nu} c_{b}\left(\bar{b}_{a} \gamma^{\nu} C \bar{b}_{b}^{T}-\bar{b}_{b} \gamma^{\nu} C \bar{b}_{a}^{T}\right) \\
\eta_{4 \mu}^{-} &=c_{a}^{T} C \gamma^{\nu} c_{b}\left(\bar{b}_{a} \sigma_{\mu \nu} C \bar{b}_{b}^{T}-\bar{b}_{b} \sigma_{\mu \nu} C \bar{b}_{a}^{T}\right)
\end{aligned}
\label{J1m}
\end{equation}
in which $\eta_{1\mu}^{-}$ and $\eta_{2\mu}^{-}$ are in symmetric color structure $[6_{c}]_{cc}\otimes[\bar{6}_{c}]_{\bar{b}\bar{b}}$ while $\eta_{3\mu}^{-}$ and $\eta_{4\mu}^{-}$ in antisymmetric color structure $[\bar{3}_{c}]_{cc}\otimes[3_{c}]_{\bar{b}\bar{b}}$.

\item The tetraquark interpolating currents with $J^{P}=1^{+}$ are
\begin{equation}
\begin{aligned}
\eta_{1 \mu}^{+} &=c_{a}^{T} C \gamma_{\mu} \gamma_{5} c_{b}\left(\bar{b}_{a} C \bar{b}_{b}^{T}+\bar{b}_{b} C \bar{b}_{a}^{T}\right) \\
\eta_{2 \mu}^{+} &=c_{a}^{T} C c_{b}\left(\bar{b}_{a} \gamma_{\mu} \gamma_{5} C \bar{b}_{b}^{T}+\bar{b}_{b} \gamma_{\mu} \gamma_{5} C \bar{b}_{a}^{T}\right) \\
\eta_{3 \mu}^{+} &=c_{a}^{T} C \sigma_{\mu \nu} \gamma_{5} c_{b}\left(\bar{b}_{a} \gamma^{\nu} C \bar{b}_{b}^{T}-\bar{b}_{b} \gamma^{\nu} C \bar{b}_{a}^{T}\right) \\
\eta_{4 \mu}^{+} &=c_{a}^{T} C \gamma^{\nu} c_{b}\left(\bar{b}_{a} \sigma_{\mu \nu} \gamma_{5} C \bar{b}_{b}^{T}-\bar{b}_{b} \sigma_{\mu \nu} \gamma_{5} C \bar{b}_{a}^{T}\right)
\end{aligned}
\label{J1p}
\end{equation}
in which $\eta_{1\mu}^{+}$ and $\eta_{2\mu}^{+}$ are in symmetric color structure $[6_{c}]_{cc}\otimes[\bar{6}_{c}]_{\bar{b}\bar{b}}$ while $\eta_{3\mu}^{+}$ and $\eta_{4\mu}^{+}$ in antisymmetric color structure $[\bar{3}_{c}]_{cc}\otimes[3_{c}]_{\bar{b}\bar{b}}$.

\item The tetraquark interpolating currents with $J^{P}=2^{+}$ are
\begin{equation}
\begin{aligned}
\eta_{1 \mu \nu}^{+}&=c_{a}^{T} C \gamma_{\mu} \gamma_{5} c_{b}\left(\bar{b}_{a} \gamma_{\nu} \gamma_{5} C \bar{b}_{b}^{T}+\bar{b}_{b} \gamma_{\nu} \gamma_{5} C \bar{b}_{a}^{T}\right) \\
\eta_{2 \mu \nu}^{+}&=c_{a}^{T} C \gamma_{\mu} c_{b}\left(\bar{b}_{a} \gamma_{\nu} C \bar{b}_{b}^{T}-\bar{b}_{b} \gamma_{\nu} C \bar{b}_{a}^{T}\right)
\end{aligned}
\label{J2p}
\end{equation}
in which $\eta_{1\mu\nu}^{+}$ is in symmetric color structure $[6_{c}]_{cc}\otimes[\bar{6}_{c}]_{\bar{b}\bar{b}}$ while $\eta_{2\mu\nu}^{+}$ in antisymmetric color structure $[\bar{3}_{c}]_{cc}\otimes[3_{c}]_{\bar{b}\bar{b}}$.

\end{itemize}

\section{QCD sum rules}
In this section, we study the two-point correlation functions induced by the interpolating tetraquark currents obtained above. For the scalar and pseudo-scalar currents, the correlation function are
\begin{equation}
\begin{aligned}
\Pi\left(p^{2}\right)&=i \int d^{4} x e^{i p \cdot x}\left\langle 0\left|T\left[J(x) J^{\dagger}(0)\right]\right| 0\right\rangle\, ,
\end{aligned}
\end{equation}
and for the vector and axial-vector currents
\begin{equation}
\begin{aligned}
 \Pi_{\mu \nu}\left(p^{2}\right) &=i \int d^{4} x e^{i p \cdot x}\left\langle 0\left|T\left[J_{\mu}(x) J_{\nu}^{\dagger}(0)\right]\right| 0\right\rangle 
\\ &=\left(\frac{p_{\mu} p_{\nu}}{p^{2}}-g_{\mu \nu}\right) \Pi_{1}\left(p^{2}\right)+\frac{p_{\mu} p_{\nu}}{p^{2}}\Pi_{0}\left(p^{2}\right)\, ,
 \label{CF_AV}
\end{aligned}
\end{equation}
in which $\Pi_{0}\left(p^{2}\right)$ and $\Pi_{1}\left(p^{2}\right)$ are the invariant functions corresponding to the spin-0 and spin-1 intermediate states, respectively. The correlation function for the tensor current is
\begin{equation}
\begin{aligned}
 \Pi_{\mu \nu,\rho \sigma}\left(p^{2}\right) &=i \int d^{4} x e^{i p \cdot x}\left\langle 0\left|T\left[J_{\mu\nu}(x) J_{\rho\sigma}^{\dagger}(0)\right]\right| 0\right\rangle
 \\
 &=\left(\eta_{\mu\rho}\eta_{\nu\sigma}+\eta_{\mu\sigma}\eta_{\nu\rho}-\frac{2}{3}\eta_{\mu\nu}\eta_{\rho\sigma}\right) \Pi_{2}\left(p^{2}\right)+\cdots \, ,
 \label{CF_T}
\end{aligned}
\end{equation}
where
\begin{equation}\
\eta_{\mu\nu}=\frac{p_{\mu} p_{\nu}}{p^{2}}-g_{\mu \nu},
\end{equation} 
and  $\Pi_{2}\left(p^{2}\right)$ is the tensor polarization function related to the spin-2 intermediate state. The $``\cdots"$ represents other structures from spin-0 and spin-1 states. We shall consider only the spin-2 component from the tensor currents in the following study. 

At the hadronic level, the correlation function can be described via the dispersion relation
\begin{equation}
\Pi\left(p^{2}\right)=\frac{\left(p^{2}\right)^{N}}{\pi} \int_{4(m_{c}+m_{b})^{2}}^{\infty} \frac{\operatorname{Im} \Pi(s)}{s^{N}\left(s-p^{2}-i \epsilon\right)} d s+\sum_{n=0}^{N-1} b_{n}\left(p^{2}\right)^{n}\, ,
\label{Cor-Spe}
\end{equation}
where the $b_n$ is the subtraction constant. In QCD sum rules, the imaginary part of the correlation function is defined as the spectral function 
\begin{equation}
\rho (s)=\frac{1}{\pi} \text{Im}\Pi(s)=f_{H}^{2}\delta(s-m_{H}^{2})+\text{QCD continuum and higher states}\, ,
\end{equation}
in which the “one pole plus continuum” parametrization is used. The parameters $f_{H}$ and $m_{H}$ are the coupling constant and mass of the lowest-lying hadronic resonance $H$ respectively 
\begin{equation}
\begin{aligned}
\langle 0|J| H\rangle &= f_{H}\, , \\
\left\langle 0\left|J_{\mu}\right| H\right\rangle &= f_{H} \epsilon_{\mu}\, , \\
\left\langle 0\left|J_{\mu\nu}\right| H\right\rangle &= f_{H} \epsilon_{\mu\nu}
 \end{aligned}
\end{equation}
with the polarization vector $\epsilon_{\mu}$ and polarization tensor $\epsilon_{\mu\nu}$.

To extract the lowest lying resonance from a particular channel, we define the moment by taking derivatives of the correlation function $\Pi(q^{2})$ in Euclidean region $Q^{2}=-q^{2}>0$
\begin{equation}
M_{n}\left(Q_{0}^{2}\right)=\left.\frac{1}{n !}\left(-\frac{d}{d Q^{2}}\right)^{n} \Pi\left(Q^{2}\right)\right|_{Q^{2}=Q_{0}^{2}}=\int_{4(m_{c}+m_{b})^{2}}^{\infty} \frac{\rho(s)}{\left(s+Q_{0}^{2}\right)^{n+1}} d s \, . 
\end{equation}
Applying the above equation to Eq.(\ref{Cor-Spe}), we can rewrite the moment as the following form
\begin{equation}
M_{n}\left(Q_{0}^{2}\right)=\frac{f_{H}^{2}}{\left(m_{H}^{2}+Q_{0}^{2}\right)^{n+1}}\left[1+\delta_{n}\left(Q_{0}^{2}\right)\right]\, ,
\end{equation}
where $\delta_{n}(Q_{0}^{2})$ represents the contributions from higher states and continuum region, and it is a function of $n$ and $Q_{0}^{2}$. Particularly, one notes that $\delta_{n}(Q_{0}^{2})$ will incline to zero as $n$ tends to infinity for a specific value of $Q_{0}^{2}$. Considering the following ratio of the moments
\begin{equation}
r\left(n, Q_{0}^{2}\right) \equiv \frac{M_{n}\left(Q_{0}^{2}\right)}{M_{n+1}\left(Q_{0}^{2}\right)}=\left(m_{H}^{2}+Q_{0}^{2}\right) \frac{1+\delta_{n}\left(Q_{0}^{2}\right)}{1+\delta_{n+1}\left(Q_{0}^{2}\right)}\, ,
\end{equation}
one may expect that $\delta_{n}(Q_{0}^{2})\approx \delta_{n+1}(Q_{0}^{2})$ when $n$ is large enough, and then the hadron mass can be extracted as 
\begin{equation}
m_{H}=\sqrt{r\left(n, Q_{0}^{2}\right)-Q_{0}^{2}}\, . \label{hadronmass}
\end{equation}

On the other hand, we can calculate the correlation function $\Pi(p^{2})$ by means of the operator product expansion (OPE) at the quark-gluon level. To evaluate the Wilson coefficients, we adopt the propagator of heavy quark in momentum space
\begin{equation}
 i S_{Q}^{a b}(p)=\frac{i \delta^{a b}}{\hat{p}-m_{Q}}
 +\frac{i}{4} g_{s} \frac{\lambda_{a b}^{n}}{2} G_{\mu \nu}^{n} \frac{\sigma^{\mu \nu}\left(\hat{p}+m_{Q}\right)+\left(\hat{p}+m_{Q}\right) \sigma^{\mu \nu}}{12}
 +\frac{i \delta^{a b}}{12}\left\langle g_{s}^{2} G G\right\rangle m_{Q} \frac{p^{2}+m_{Q} \hat{p}}{(p^{2}-m_{Q}^{2})^{4}}\, , 
\end{equation}
where $Q$ denotes the charm or bottom quark. The superscripts $a, b$ represent the color indices and $\hat{p}=p^{\mu}\gamma_{\mu}$. In this work, we will calculate the perturbative term and gluon condensate term in the correlation function. The contributions from non-perturbative  terms with higher dimension, such as the tri-gluon condensate, are small enough to be neglected.

\section{Numerical analysis}
In this section, we perform the QCD moment sum rule analyses for the $cc\bar{b}\bar{b}$ states. We use the values of heavy quark masses and gluon condensate as follows~\cite{Nielsen:2009uh,Narison:2018nbv,Zyla:2020zbs}
\begin{equation}
\begin{array}{l}
{m_{c}\left(m_{c}\right)=(1.27 \pm 0.02) ~\mathrm{GeV}}\, , \vspace{1ex} \\
{m_{b}\left(m_{b}\right)=(4.18 _{-0.02}^{+0.03})~ \mathrm{GeV}}\, , \vspace{1ex} \\
{\left\langle g_{s}^{2} G G\right\rangle= (0.88\pm0.25)~ \mathrm{GeV}^{4}}\, .
\end{array}
\end{equation}
Besides, we consider the scale dependence of the charm and bottom quark masses at the leading order
\begin{equation}
\begin{aligned} m_{c}(\mu) &=\bar{m}_{c}\left(\frac{\alpha_{s}(\mu)}{\alpha_{s}\left(\bar{m}_{c}\right)}\right)^{12 / 25}\, ,
 \\ m_{b}(\mu) &=\bar{m}_{b}\left(\frac{\alpha_{s}(\mu)}{\alpha_{s}\left(\bar{m}_{b}\right)}\right)^{12 / 23}\, , 
 \end{aligned}
\end{equation}
where the evolution of strong coupling 
\begin{equation}
\alpha_{s}(\mu)=\frac{\alpha_{s}\left(M_{\tau}\right)}{1+\frac{25 \alpha_{s}\left(M_{\tau}\right)}{12 \pi} \log \left(\frac{\mu^{2}}{M_{\tau}^{2}}\right)}, \quad \alpha_{s}\left(M_{\tau}\right)=0.33
\end{equation}
is applied, and the $\tau$ mass $M_{\tau}$  is  uesd from PDG values~\cite{Zyla:2020zbs}. For $cc\bar{b}\bar{b}$ system, we use the renormalization scale $\mu=\frac{\bar{m}_{c}+\bar{m}_{b}}{2}=2.73 \mathrm{GeV}$ in our moment sum rule analysis.

In Eq.~\eqref{hadronmass}, there are two parameters $n$ and $Q_{0}^{2}$ for the hadron mass prediction. A suitable working region for these two parameters $(n,\, Q_{0}^{2})$ is needed to obtain stable and reliable mass sum rules.
In Ref.~\cite{Shifman:1978bx,Shifman:1978by}, the parameter $Q_{0}^{2}=0$ was adopted in the moment sum rules, which leads to a bad convergence of OPE series. In this paper, we shall follow Ref.~\cite{Chen:2016jxd} to choose $Q_{0}^{2}>0$ and define $\xi=Q_{0}^{2}/(m_{b}+m_{c})^{2}$ to perform mass sum rule analysis to avoid the bad OPE convergence problem. As matter of fact, the selections of the parameters $n$ and $\xi$ are restricted from the following two prospects: (1)~$n$ should be large enough to reduce the contributions from higher states and continuum region, but it will also decrease the convergence of OPE. (2) a large $\xi$ (or $Q_{0}^{2}$) will also reduce the convergence of $\delta_{n}(Q_{0}^{2})$ which makes it difficult to extract the parameters of the lowest lying resonance. 

We take the interpolating current $\eta_{1}^{-}$ with $J^{P}=0^{-}$ as an example to show the details of our numerical analysis. The correlation function for the current is calculated including only perturbative term and gluon condensate 
\begin{equation}
\begin{aligned} 
\Pi^{pert}(Q^{2})&= \frac{1}{32\pi^{6}}\int_{0}^{1}dx\int_{0}^{1}dy\int_{0}^{1}dz
\Bigg\{\left[ \frac{-3x(1-x)y(1-y)^{3}(1-z)}{z^{3}}\right]F(m_{c},m_{b},Q^{2})^{4} \\
&+\left[\frac{4m_{b}^{2}y(1-y)}{z^{2}}-\frac{4m_{c}^{2}x(1-x)(1-y)^{3}(1-z)}{z^{3}} -\frac{12Q^{2}x(1-x)y(1-y)^{3}(1-z)^{2}}{z^{2}}\right]F(m_{c},m_{b},Q^{2})^{3}\\
&+6\left[\frac{m_{b}^{2}m_{c}^{2}(1-y)}{z^{2}}+\frac{m_{b}^{2}Q^{2}y(1-y)(1-z)}{z}-\frac{m_{c}^{2}Q^{2}x(1-x)(1-y)^{3}(1-z)^{2}}{z^{2}}\right.\\
&\left. -\frac{Q^{4}x(1-x)y(1-y)^{3}(1-z)^{3}}{z}\right]F(m_{c},m_{b},Q^{2})^{2}\Bigg\}Log[F(m_{c},m_{b},Q^{2})]\\
\Pi^{GG}(Q^{2})&= \frac{\langle g_{s}^{2}GG \rangle}{96\pi^{6}}\int_{0}^{1}dx\int_{0}^{1}dy\int_{0}^{1}dz
\Bigg\{\left[ 6 \left(\frac{m_{b}^{2} y}{x^{2} (1-y)}-\frac{m_{b}^{2} (1-x) y (1-z)}{x^{2}}\right)\right.\\
&\left. +6 \left(-\frac{m_{c}^{2} x (1-x) (1-y)^{3} (1-z)^{3}}{z^{3}}-\frac{m_{c}^{2} x (1-x) y (1-y)^{3} (1-z)^{4}}{z^{3}}\right)\right]F(m_{c},m_{b},Q^{2}) \\
&+\left(\frac{2 m_{b}^{2} m_{c}^{2} y (1-y) (1-z)^{3}}{z^{2}}+\frac{3 m_{b}^{2} m_{c}^{2} (1-y) (1-z)^{2}}{z^{2}}
-\frac{2 m_{c}^{4} (1-x) x (1-y)^{3} (1-z)^{4}}{z^{3}}\right.\\
&\left. -\frac{6 m_{c}^{2} Q^{2} x (1-x) y (1-y)^{3} (1-z)^{5}}{z^{2}}-\frac{3 m_{c}^{2} Q^{2}(1-x) x (1-y)^{3} (1-z)^{4}}{z^{2}}\right)\\
&+\left(\frac{2 m_{b}^{4} y z}{x^{3} (1-y)^{2}}+\frac{3 m_{b}^{2} m_{c}^{2}}{x^{2} (1-y)}-\frac{2 m_{b}^{2} m_{c}^{2} (1-x) (1-z)}{x^{2}}-\frac{6 m_{b}^{2} Q^{2}(1-x) y z (1-z)^{2}}{x^{2}}\right.\\
&+\left.\frac{3 m_{b}^{2} Q^{2}y z (1-z)}{x^{2} (1-y)}\right)\Bigg\}Log[F(m_{c},m_{b},Q^{2})]\\
&+\frac{\langle g_{s}^{2}GG \rangle}{96\pi^{6}}\int_{0}^{1}dx\int_{0}^{1}dy\int_{0}^{1}dz \frac{1}{F(m_{c},m_{b},Q^{2})}
\Bigg\{
\left(\frac{m_{b}^{2}m_{c}^{4}  (1-y) (1-z)^{3}}{z^{2}}+\frac{m_{b}^{2}m_{c}^{2}Q^{2}y (1-y) (1-z)^{4}}{z}\right.\\
&\left.-\frac{m_{c}^{4} Q^{2}(1-x) x (1-y)^{3} (1-z)^{5}}{z^{2}}-\frac{m_{c}^{2}Q^{4} x (1-x) y (1-y)^{3} (1-z)^{6}}{z}\right)+\left(\frac{m_{b}^{4}  m_{c}^{2}z}{x^{3} (1-y)^{2}}\right.\\
&\left.+\frac{m_{b}^{4} Q^{2}y z^{2} (1-z)}{x^{3} (1-y)^{2}}-\frac{m_{b}^{2}m_{c}^{2}Q^{2}(1-x)z(1-z)^{2} }{x^{2}}-\frac{m_{b}^{2}Q^{4} (1-x) y z^{2} (1-z)^{3}}{x^{2}}\right)\Bigg\}\\
&+\frac{\langle g_{s}^{2}GG \rangle}{256\pi^{6}}\int_{0}^{1}dx\int_{0}^{1}dy\int_{0}^{1}dz \Bigg\{\left[\frac{3 (1-x) x (1-y)^{3} (1-z)^{2}}{z^{2}}+\frac{3 y (1-y) (1-z)}{z}\right]F(m_{c},m_{b},Q^{2})^{2}\\
&+\left[\left(-\frac{2 m_{b}^{2} (1-y) (1-z)}{z}+\frac{4 m_{c}^{2} x (1-x) (1-y)^{3} (1-z)^{2}}{y z^{2}}+\frac{6 Q^{2}(1-x) x (1-y)^{3} (1-z)^{3}}{z}\right)\right.\\
&\left.+\left(\frac{2 m_{c}^{2} (1-y) (1-z)}{z}+6 Q^{2}(1-y) y (1-z)^{2}-\frac{4 m_{b}^{2} y}{x (1-x) (1-y)}\right)\right]F(m_{c},m_{b},Q^{2})\\
&+\left(\frac{2 m_{c}^{2} Q^{2}x (1-x) (1-y)^{3} (1-z)^{3}}{y z}-m_{b}^{2} Q^{2}(1-y) (1-z)^{2}+Q^{4} (1-x) x (1-y)^{3} (1-z)^{4}\right.\\ &\left.-\frac{2 m_{b}^{2} m_{c}^{2} (1-y) (1-z)}{y z}\right)
+\left(-\frac{2 m_{b}^{2} m_{c}^{2}}{x (1-x) (1-y)}+m_{c}^{2} Q^{2}(1-y) (1-z)^{2}\right.\\ &\left.+Q^{4} (1-y) y z (1-z)^{3}-\frac{2 m_{b}^{2} Q^{2}y z (1-z)}{x (1-x) (1-y)}\right)\Bigg\}Log[F(m_{c},m_{b},Q^{2})]\, ,
\end{aligned}
\end{equation}
where $F(m_{c},m_{b},Q^{2})=m_{c}^{2}(1-z)+\frac{m_{c}^{2} z}{y}+\frac{m_{b}^{2} z}{x(1-y)}+\frac{m_{b}^{2} z}{(1-x)(1-y)}+Q^{2} z(1-z)$. We don't list all the correlation function expressions for other interpolating currents in Eqs.~\eqref{J0m}-\eqref{J2p} since they are very lengthy to be shown here.

\begin{figure}[t!]
\centering
\includegraphics[width=10cm]{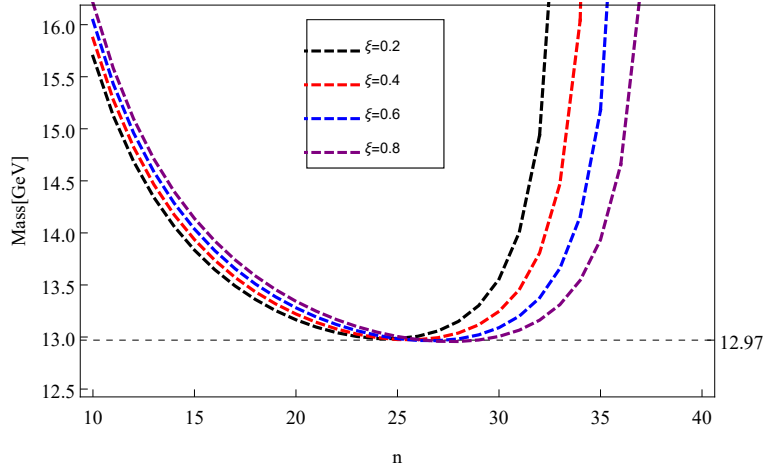}\\
\caption{Hadron mass of $cc\bar{b}\bar{b}$ with $J^{P}=0^{-}$ from $\eta_{1}^{-}$ with respect to $n$ for different $\xi$.}
\label{massVSn}
\end{figure}

\begin{figure}[t!]
\centering
\includegraphics[width=10cm]{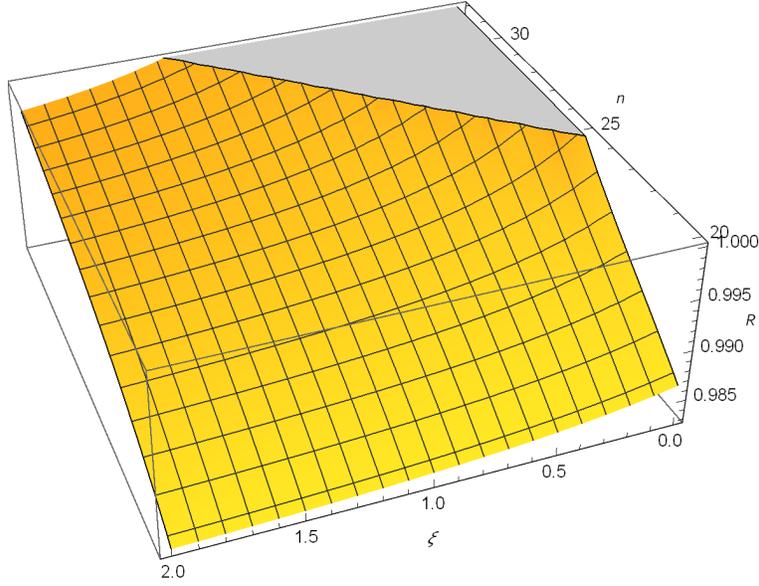}\\
\caption{Ratio $R$ as a function of $n$ and $\xi$.}
\label{n_R_xi}
\end{figure}

The upper bound of the parameter $n$ can be determined by ensuring the OPE convergence. We here require that the contribution from perturbative term to be larger than that from the gluon condensate term, and then the upper bound of parameter $n$ can be obtained as $n_{max}=36, 37, 39, 40$ for $\xi=0.2, 0.4, 0.6, 0.8$ respectively. In Fig.~(\ref{massVSn}), we show the curves of the extracted mass versus $n$ with different $\xi$. The stable mass prediction plateau can be found where the variations of hadron mass with respect to $n$ and $\xi$ minimize. We find that the same mass prediction plateaus can also be obtained in the following way, i.e, the Schwarz inequality should be satisfied by the moment $M_{n}\left(Q_{0}^{2}\right)$
\begin{equation}
R=\frac{M_{n}\left(Q_{0}^{2}\right)^{2}}{M_{r}\left(Q_{0}^{2}\right) M_{2 n-r}\left(Q_{0}^{2}\right)} \leq 1 \, ,
\end{equation}
where $r<2n$. We show the variations of ratio $R$ with respect to $n$ and $\xi$ in Fig.(\ref{n_R_xi}) where the gray part represents $R>1$ region while the yellow part denotes $R<1$ region.  The dividing line between the two parts lead to the values of $(n, \xi)=(25, 0.2), (26, 0.4), (27, 0.6), (28, 0.8)$, which can also be obtained at the plateaus in the mass prediction curves. One can see that this plateaus
provide a much stronger constrain for the $(n, \xi)$ plane than requiring the convergence of OPE series. Then the extracted mass from $\eta_{1}^{-}$ with $J^{P}=0^{-}$ is
\begin{equation}
m_{1}=12.97_{-0.21}^{+0.25} ~\text{GeV}\, ,
\end{equation}
where the errors are from the uncertainties of $\xi$ and $n$, heavy quark masses and the gluon condensate.

By performing similar analyses to interpolating currents in Eqs.~\eqref{J0m}-\eqref{J2p}, we can find the stable mass sum rules for all channels and collect the predicted masses of these $cc\bar b\bar b$ tetraquark states in Table~\ref{ccbbResultTab}. It shows that the masses for $J^{P}=0^{+}$($\eta_{2}^+$ $\sim$ $\eta_{5}^+$), $1^+$ ($\eta_{3\mu}^+$ and $\eta_{4\mu}^+$) and $2^{+}$($\eta_{2\mu\nu}$) $cc\bar{b}\bar{b}$ tetraquark states are about $12.3-12.4$ GeV, while the masses for $J^{P}=0^{-}$ and $1^{-}$ $cc\bar{b}\bar{b}$ tetraquarks are about $12.8-13.1$ GeV.  One notes that the mass predictions for $\eta_{1}^{+}$ with $J^{P}=0^{+}$, $\eta_{1\mu}^{+}$ and $\eta_{2\mu}^{+}$ with $J^{P}=1^{+}$ and $\eta_{1\mu\nu}^{+}$ with $J^{P}=2^{+}$, are much higher than those from the currents 
with the same corresponding quantum numbers respectively, which may be originating from the two P-wave diquark-antidiquark pair in the currents 
$\eta_{1}^{+}$, $\eta_{1\mu}^{+}$, $\eta_{2\mu}^{+}$ and $\eta_{1\mu\nu}^{+}$.

\begin{table*}[h!]
\caption{The hadron mass predictions for the $cc\bar{b}\bar{b}$ tetraquark states with various $J^{P}$.}
\renewcommand\arraystretch{1.8} 
\setlength{\tabcolsep}{2.em}{ 
\begin{tabular}{ccc|ccc}
 \hline  \hline      
   Current             & $J^{P} $     &Mass(\text{GeV)}             &Current          &$J^{P} $   & Mass(\text{GeV)}  \\ \hline
    $\eta_{1}^{+}$     & $0^{+}$     & $13.32_{-0.24}^{+0.30} $    & $\eta_{1}^{-}$      & $0^{-}$      &  $12.97_{-0.21}^{+0.25} $  \\              
   $\eta_{2}^{+}$      & $0^{+}$     & $12.41 _{-0.17}^{+0.21} $    & $\eta_{2}^{-}$      & $0^{-}$      &  $12.72_{-0.19}^{+0.22} $   \\

   $\eta_{3}^{+}$      & $0^{+}$     & $12.33 _{-0.15}^{+0.18} $    & $\eta_{3}^{-}$      & $0^{-}$      &  $13.16_{-0.24}^{+0.29} $   \\

   $\eta_{4}^{+}$      & $0^{+}$     & $12.36 _{-0.15}^{+0.18} $    & & &  \\  
   $\eta_{5}^{+}$      & $0^{+}$     & $12.36 _{-0.16}^{+0.19} $   &   &  &   \\ [.2 cm]  
    $\eta_{1\mu}^{+}$  & $1^{+}$     & $13.35_{-0.26}^{+0.33} $     & $\eta_{1\mu}^{-}$   & $1^{-}$      &  $13.02_{-0.21}^{+0.26} $   \\ 
    $\eta_{2\mu}^{+}$  & $1^{+}$     & $13.33_{-0.22}^{+0.28} $     & $\eta_{2\mu}^{-}$   & $1^{-}$      &  $12.77_{-0.19}^{+0.24} $     \\
    $\eta_{3\mu}^{+}$  & $1^{+}$     & $12.36_{-0.16}^{+0.19} $     & $\eta_{3\mu}^{-}$   & $1^{-}$      &  $12.99_{-0.22}^{+0.27} $    \\
    $\eta_{4\mu}^{+}$  & $1^{+}$     & $12.34_{-0.15}^{+0.18} $     & $\eta_{4\mu}^{-}$   & $1^{-}$      &  $12.87_{-0.20}^{+0.24} $     \\ [.2 cm]  
     $\eta_{1\mu\nu}^{+}$     & $2^{+}$                   &  $13.41_{-0.26}^{+0.34} $   &&&    \\
     $\eta_{2\mu\nu}^{+}$     & $2^{+}$                   &  $12.37_{-0.16}^{+0.19} $   &&&    \\
       \hline\hline  
\label{ccbbResultTab}
\end{tabular}
}
\end{table*}



\section{Conclusion and Discussion}
We have investigated the mass spectra for the $cc\bar{b}\bar{b}$ tetraquark states in the framework of QCD moment sum rules. We construct the interpolating tetraquark currents with $J^{P}=0^{\pm}$, $1^{\pm}$and $2^{+}$ and calculate their two-point correlation functions containing perturbative term and gluon condensate term. Performing the QCD moment sum rules, we obtain stable sum rules for all currents.

Our results show that the masses for the positive parity $cc\bar{b}\bar{b}$ tetraquarks with $J^{P}=0^{+}$($\eta_{2}^+$ $\sim$ $\eta_{5}^+$), $1^+$ ($\eta_{3\mu}^+$ and $\eta_{4\mu}^+$) and $2^{+}$($\eta_{2\mu\nu}$) $cc\bar{b}\bar{b}$ tetraquark states are about $12.3-12.4$ GeV, and the masses for the negative parity tetraquarks with $J^{P}=0^{-}$ and $1^{-}$ $cc\bar{b}\bar{b}$ are $12.8-13.1$ GeV. Such mass difference is reasonable since that the positive parity channels are S-wave tetraquarks and the negative parity channels are P-wave ones,  Noting that 
the diquarks and antidiquarks in the currents 
$\eta_{1}^{+}$, $\eta_{1\mu}^{+}$, $\eta_{2\mu}^{+}$ and $\eta_{1\mu\nu}^{+}$ are all $P$-wave operators, their masses are predicted to be much higher than those from the other positive parity currents, in which the diquarks and antidiquarks are all in S-wave. 

Carrying double different flavors, the $cc\bar{b}\bar{b}$ tetraquark states can not decay into a heavy quarkonium plus a light meson via annihilating a pair of heavy quark-antiquark. There is only one kind of two-meson strong decay threshold $B_{c}^{(*)}B_{c}^{(*)}$ for the $cc\bar{b}\bar{b}$ tetraquark systems. To date, only one ground state of $B_c$ meson has been discovered and confirmed experimentally with $m_{B_{c}}=6.25$ GeV~\cite{Zyla:2020zbs,1998-Abe-p112004-112004}. 
The spectroscopy of the other $B_c$ mesons has been calculated in the relativistic quark model~\cite{2004-Godfrey-p54017-54017}, in which the 
mass of  $B_{c}^{*}$ with $J^{P}=1^{-}$ was predicted to be $m_{B_{c}^{*}}=6.34$ GeV.


In Table~\ref{ccbbResultTab}, the negative parity $cc\bar{b}\bar{b}$ tetraquarks with $J^{P}=0^{-}$ and $1^{-}$ are predicted to be above the $B_{c}B_{c}^{*}$ and $B_{c}^{*}B_{c}^{*}$ 
thresholds and thus can decay into these final state via strong interaction. Besides, the masses of doubly P-wave tetraquarkes from currents 
$\eta_{1}^{+}$, $\eta_{1\mu}^{+}$, $\eta_{2\mu}^{+}$ and $\eta_{1\mu\nu}^{+}$ are above $2B_{c}^{*}$ threshold, and can decay into $B_{c} B_{c}$, $B_{c} B_{c}^{*}$, $B_{c}^{*}B_{c}^{*}$. However, the positive parity $cc\bar{b}\bar{b}$ tetraquarks with  $J^{P}=0^{+}$($\eta_{2}^+$ $\sim$ $\eta_{5}^+$), $1^+$ ($\eta_{3\mu}^+$ and $\eta_{4\mu}^+$) and $2^{+}$($\eta_{2\mu\nu}$) are predicted to be below the $2B_c$ threshold, implying that these tetraquark states can only undergo radiative transitions or weak decays. They are expected to be very narrow and stable if they do exist.

\section*{ACKNOWLEDGMENTS}

This project is supported by the National Key Research and Development Program of China (2020YFA0406400), the National Natural Science Foundation of China under Grants No. 11722540 and No. 12075019, the Fundamental Research Funds for the Central Universities.


\end{document}